# Unexpectedly large entropic barrier controls bond rearrangements in vitrimers

*Shinian Cheng[1*], Lilliana Rey[1], Beibei Yao[2], Ivan Popov[3], Alexei P. Sokolov[1,2*]*

[1]*Department of Chemistry, University of Tennessee, Knoxville, Tennessee 37996, USA*

[2]*Chemical Sciences Division, Oak Ridge National Laboratory, Oak Ridge, Tennessee 37831, USA*

[3]*University of Tennessee-Oak Ridge Innovation Institute, University of Tennessee, Knoxville, Tennessee 37996, USA*

**Abstract:** Vitrimers are a relatively new class of polymeric materials containing associative covalent dynamic bonds that make them recyclable by design. However, the fundamental mechanisms controlling their viscoelastic properties remain poorly understood. Our detailed studies of relaxation dynamics and viscoelastic behavior of model vitrimers revealed that the density of dynamic covalent crosslinks has no influence on chain dynamics (beyond a weak change in the glass transition temperature), yet it strongly affects the linear viscoelasticity of vitrimers. Increasing the crosslink density induces a sol-gel transition consistent with predictions of classical gelation theory, demonstrating its applicability to vitrimers. Remarkably, the temperature-dependent analysis of the bond rearrangement time reveals an unexpectedly large negative activation entropy in the transition state that strongly slows down the bond exchange process despite its relatively low activation enthalpy. This insight explains the unusual long timescale for bond rearrangement in vitrimers and highlights the significance of entropy in controlling the viscoelasticity of dynamic covalent networks.

* Corresponding Authors. scheng17@utk.edu; sokolov@utk.edu

Vitrimers represent a unique class of polymer materials characterized by dynamic covalent networks formed through associative bond exchange reactions [1-3]. These materials can show many advanced functionalities, including recyclability, self-healing, shape memory, and extreme toughness [4-12], making them promising candidates for next-generation functional materials [13-15]. However, due to their reversible nature, dynamic covalent bonds generate additional parameters controlling the viscoelastic properties and new relaxation processes in vitrimers. This significantly complicates the fundamental understanding of the mechanisms controlling viscoelasticity and relaxation dynamics of vitrimers, which presents a major challenge in the rational design of vitrimers with tailored properties such as reduced creep, improved recyclability and self-healing.

A central question in the field of vitrimers is what controls the bond rearrangement time $\tau_r$, particularly, its temperature-dependence. In classical models of reversible networks, $\tau_r$ is often described by an Arrhenius-type relationship [16-18]:

$$\tau_r(T) = \tau_x(T) exp\left(\frac{E_a}{k_B T}\right) \approx B\tau_\alpha(T) exp\left(\frac{E_a}{k_B T}\right) \qquad (1)$$

Here, $k_B$ is the Boltzmann constant, $T$ is the absolute temperature, $E_a$ is the energy barrier for bond rearrangement, and the pre-factor $\tau_x(T)$ is traditionally assumed to follow the segmental relaxation time $\tau_\alpha(T)$ with some coefficient $B$, reflecting reversible bonds diffusion and their potential multiple returns to the same bond before finding a new partner [16,19]. However, studies of dynamic covalent networks revealed that $\tau_r$ can exhibit a much weaker temperature dependence than $\tau_\alpha$ [20,21], which contradicts to Eq. (1). This suggests that segmental dynamics plays a limited role in bond rearrangement under the temperature regime of interest. It is also possible that chain dynamics controlling stickers' diffusion have weaker temperature dependence than segmental relaxation, as it is known from the failure of time-temperature superposition in many polymers [22-27]. Unfortunately, chain dynamics were not studied as extensively as segmental dynamics [28], leaving open questions regarding the decoupling between segmental and chain dynamics in vitrimers. To explain the weaker temperature dependence of $\tau_r$, one suggests a dependence on the sum of two contributions: the bond exchange kinetics determined by chemical reaction and the stickers diffusion controlled by segmental dynamics [29,30]:

$$\tau_r(T) = A \exp\left(\frac{E_a}{k_B T}\right) + B\tau_\alpha(T) \qquad (2)$$

Here $A$ is the pre-factor. At high temperatures well above the glass transition temperature $T_g$ where stickers diffusion is rapid, Eq. (2) simplifies to an Arrhenius-type law:

$$\tau_r(T) = A\exp\left(\frac{E_a}{k_B T}\right) \qquad (3)$$

Recent studies [31,32] have shown that Eq. (3) effectively describes $\tau_r(T)$ at high temperatures and gives reasonably consistent activation energies for bond exchange. Interestingly, these studies have also revealed an extremely large pre-factor $A$ in vitrimers bearing boronic/boric esters (with $A \sim 10^{-2}$-$10^{-7}$ s) and imine bonds (with $A \sim 10^{-2}$-$10^{-10}$ s), which are considerably larger than the typical elementary timescale of pre-factor $\sim 10^{-12}$-$10^{-13}$ s [33,34]. Carden *et. al* [32] ascribed these large pre-factors to the chemical steric factor, which defines the probability of successful collision geometry resulting in a chemical reaction [35]. Nevertheless, it is still unclear what controls the steric factor, and therefore, a satisfactory physical description of the bond rearrangement time in vitrimers has not been achieved yet.

Another unresolved issue is the sol-gel transition in dynamic covalent networks. The classical Flory-Stockmayer theory [36,37] describes gelation as a percolation process involving formation of permanent crosslinks. However, crosslinks in vitrimers are dynamic with finite bond exchange time. This raises the fundamental questions on whether a "gel point" in the classical sense can be clearly defined or whether it becomes a kinetically shifting threshold governed by a balance between bond formation/exchange and polymer mobility. Although relaxation dynamics in vitrimers well above gelation have been extensively studied, the fundamental thermodynamics and dynamics near the gelation point remain largely unexplored.

To address these issues, we studied the relaxation dynamics and viscoelastic properties of poly(propylene glycol) (PPG)-based vitrimers with various crosslink densities. We found that the density of dynamic covalent crosslinks has almost no impact on segmental and chain dynamics, however, it significantly influences the linear viscoelastic behavior of studied samples. Increasing crosslink density induces a sol-gel transition with the gelation point being consistent with predictions of classical sol-gel theory for permanent networks, revealing its applicability to vitrimers. Surprisingly, the temperature dependence of bond rearrangement times reveals unusually large negative activation entropies. This finding explains the unexpectedly slow bond rearrangement process in vitrimers with low enthalpic barriers and reveals the critical role of entropy in governing the viscoelasticity of vitrimers.

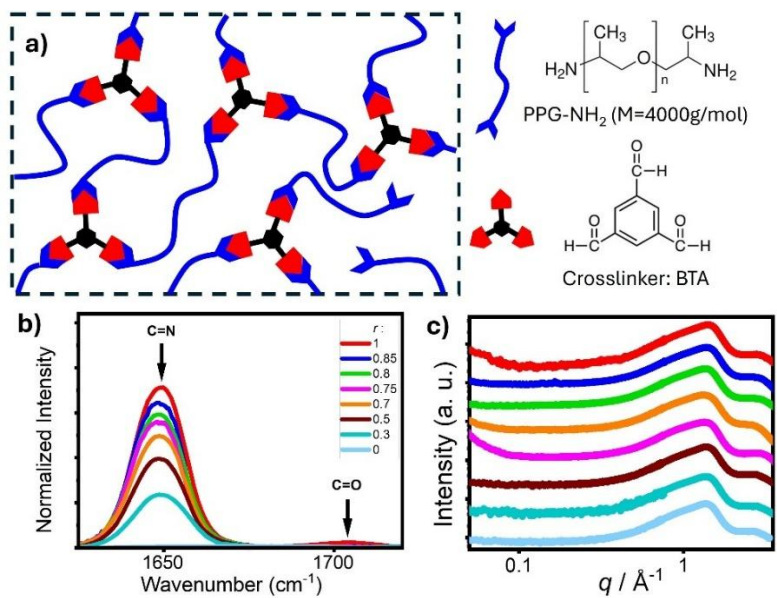

**Fig. 1 Molecular design and structure of vitrimers**. a) A schematic of the studied vitrimer network bearing imine bonds. b) Raman spectra confirming the formation of imine bonds and quantifying the imine bond concentrations. c) X-ray scattering data confirms absence of phase separation in all synthesized samples.

We designed model vitrimers bearing imine bonds using $NH_2$-terminated telechelic PPG (molecular weight $M$=4000 g/mol) crosslinked by benzene-1,3,5-tricarbaldehyde (BTA) (Fig.1a). The advantages of this vitrimer are: 1) the molecular weight between two stickers is constant; 2) the polymer chain dynamics can be explored using broadband dielectric spectroscopy (BDS) due to the accumulated diploe along the PPG backbone; therefore, it provides the possibility to explore the potential role of polymer chain dynamics in bond rearrangements of vitrimers. The synthesized samples are named PPG$r$, where $r$ denotes the molar ratio between -CHO groups in BTA and -$NH_2$ groups in PPG. For example, PPG0.8 means that the molar ratio of -CHO to -$NH_2$ for the reaction is 0.8 and, consequently, the expected fraction of -$NH_2$ conversion to imine bonds is 0.8. Moreover, the polymer molecular weight 4000 g/mol is below the critical entanglement molecular weight of poly(propylene glycol), $M_c \sim$ 5300-7000 g/mol [38,39], so the effects of entanglements are negligible. Raman spectrometer (JY Horiba 64000) and small angle X-ray scattering (Xenocs

Xeuss 3.0) were employed to verify the chemical structures and the absent of nanoscale phase separation of synthesized samples, respectively. Dielectric spectra in the frequency range between $10^{-2}$ and $10^6$ Hz were measured using a Novocontrol system to investigate the segmental and chain dynamics. The small amplitude oscillatory shear (SAOS) rheology measurements were conducted in the linear regime with the strain-controlled mode of an AR2000ex rheometer in the angular frequency range of $10^{-2}$-$10^2$ rad/s. Detailed information about the synthesis and experimental procedure can be found in the Supplementary Materials (SM) [40].

Raman spectroscopy confirmed and quantified the formation of imine bonds in synthesized samples. Compared to the polymer precursor (*i.e.*, PPG0), the Raman spectra of synthesized samples show an additional peak at 1649 cm$^{-1}$, characterizing the C=N bond stretching (Figure 1b). Quantitatively, the peak area at 1649 cm$^{-1}$ increases with an increase in crosslinker density. These spectra were normalized to the peak at 1099 cm$^{-1}$, characterizing the C-O bonds in the polymer backbone (Fig. S1). Notably, a tiny peak at 1705 cm$^{-1}$ (characterizing C=O bond) was observed only in PPG1 since a slight excess of BTA was added during its synthesis to make sure no free amine groups remained. This peak is absent in the spectra of other samples indicating complete conversion of C=O bonds of BTA to C=N crosslinks. Small angle X-ray scattering demonstrated the absence of nanoscale phase separation in all samples (Fig. 1c). Furthermore, the differential scanning calorimetry (DSC) measurement (Fig. S2) reveals a single glass transition with $T_g$ increasing systematically with crosslinks density, which also confirms the absence of micro-phase separation.

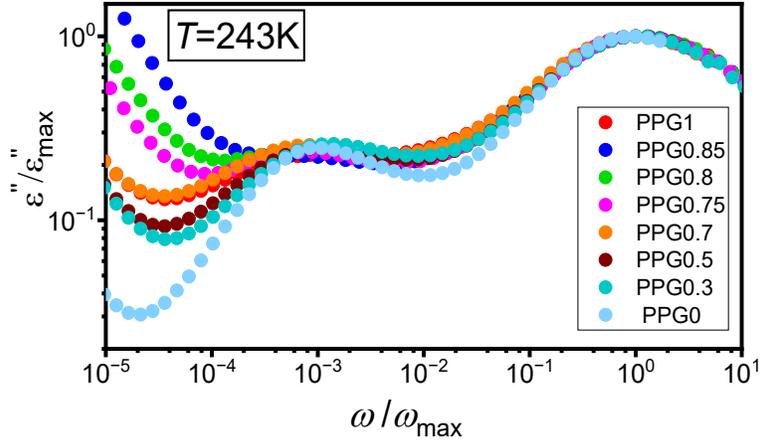

**Fig. 2** Imaginary part of the dielectric permittivity spectra in studied samples scaled at the segmental relaxation peak.

Two main relaxation processes are observed in the BDS spectra (Figs. 2 and S3), with the higher frequency peak assigned to segmental relaxation process and the lower frequency one assigned to the normal mode (chain relaxation) process [16], reflecting the fluctuations of end-to-end dipoles of the polymer chain [41]. To account for the small changes in the glass transition temperature, $T_g$ (Fig. S2), and in the segmental relaxation time, we normalized $\varepsilon''(\omega)$ to the maximum of the segmental relaxation peak (Fig. 2). Surprisingly, such normalization of BDS spectra (Fig. 2) revealed that the chain relaxation dynamics remain unchanged after scaling out weak change in $T_g$, and the normal mode only broadens with increasing crosslinker concentrations. To extract the characteristic times of segmental and normal mode relaxation, the dielectric spectra were fit to two Havriliak-Negami (HN) functions (see SM for details) [42]. The obtained segmental ($\tau_s$) and normal mode ($\tau_n$) relaxation times show no dependence on crosslink density when plotted vs. $T_g/T$ (Fig. 3).

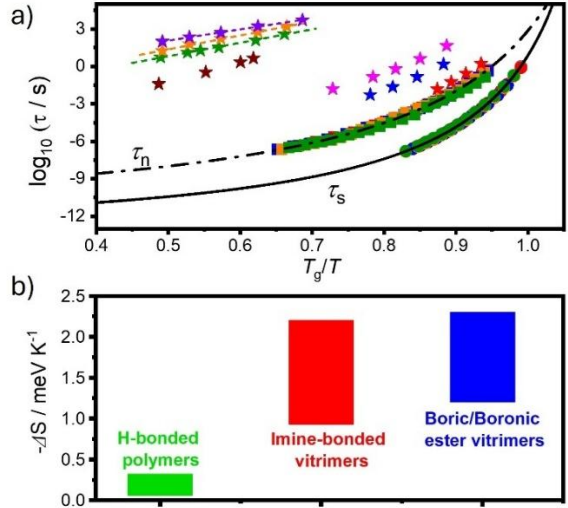

**Fig. 3.** a) Characteristic timescale for segmental relaxation ($\tau_s$, circles), normal mode ($\tau_n$, squares), and terminal flow ($\tau_f$, stars)/bond rearrangement ($\tau_r$, stars). The black solid and dotted-dashed line denotes the fit of $\tau_s$ and $\tau_n$ to the Vogel-Fulcher-Tammann (VFT) equation. Dashed lines denote the fit of $\tau_r$ to Eq. 3. b) Illustration of the activation entropy for bond rearrangement process in H-bonded associative polymers and vitrimers bearing imine bonds and boric/boronic ester bonds.

The time-temperature superposition (TTS) was employed to construct the master curve of $G'(\omega)$ and $G''(\omega)$ at the reference temperature $T_{\text{ref}}$=230K (Figure S4). Although the TTS fails to perfectly bridge the whole experimental window for the studied samples, the constructed master curves still provide valuable qualitative information on the rheological behavior of the samples. To account for the slight difference in segmental relaxation time, the obtained master curves (Fig. S4) for $G'(\omega)$ and $G''(\omega)$ were normalized at the segmental relaxation mode (Figs. 4a and 4b). The bond rearrangement time of PPG 0.85/0.8/0.75/0.7 was obtained as $\tau_r = 1/\omega_{cross}$, where $\omega_{cross}$ is the angular frequency of the $G'(\omega)$ and $G''(\omega)$ crossover at the lowest frequency side (Fig. S4). Since there is no $G'(\omega)$ and $G''(\omega)$ crossover at the terminal flow regime of PPG0.5/0.3/0 (Fig. S4), the terminal flow times of these samples were determined as $\tau_f = 1/\omega_f$ where $\omega_f$ is the frequency when the scaling laws of $G'(\omega) \propto \omega^2$ and $G''(\omega) \propto \omega$ begin. To avoid problems with the failure of TTS, $\tau_r$ and $\tau_f$ (Fig. 3a) were estimated only at temperatures when this process is in the frequency range of the rheometer.

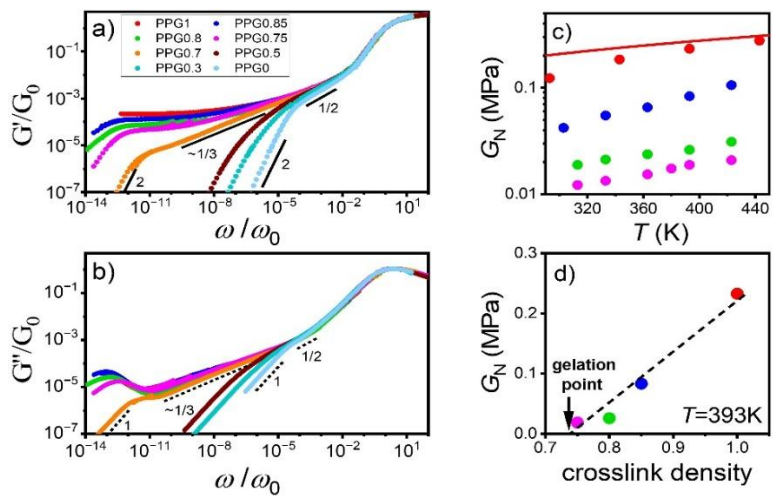

**Fig. 4.** a) Master curve of $G'$ for studied systems normalized at the segmental relaxation process. b) Master curve of $G''$ for studied systems normalized at the segmental relaxation process. c) Rubbery modulus $G_N$ plotted as a function of temperature for PPG0.75 (pink), PPG0.8 (green), PPG0.85 (blue), and PPG1 (red). The solid red line shows predictions of the phantom network model for PPG1. d) Rubbery modulus $G_N$ as a function of crosslinker density at $T$=393K. The dashed black lines are linear extrapolation to estimate the gelation point.

Crosslinking density plays a significant role in the linear viscoelastic properties of the studied samples. The increase in crosslinker concentration leads to three main changes in the linear viscoelastic spectra (Figs. 4a, 4b and S4): 1) a significant shift in the terminal flow/bond rearrangement regime to a lower frequency, highlighting the slowdown of terminal flow/bond rearrangement times; 2) the appearance of an extended intermediate power law regime of $G^*(\omega) \sim \omega^{1/3}$ for PPG0.7; and 3) the appearance of a clear rubbery plateau when crosslinker concentration is above 0.7, demonstrating the formation of a crosslinked network.

Flory-Stockmayer theory [36,37] predicts that for a system with chains $A_2$ (functionality 2, *e.g.*, PPG-NH$_2$) and crosslinkers $B_3$ (functionality 3, *e.g.* BTA), the criterion for gelation obeys $p_A \times p_B > \frac{1}{(2-1)\times(3-1)} = \frac{1}{2}$, where $p_A$ and $p_B$ are the conversions of $A$ and $B$ groups, respectively.

For balanced systems where the number of converted $A$ groups is equal to that of $B$ groups, we have $p_A = p_B$ and therefore the critical conversion for gelation is $p_{A,C} = p_{B,C} = \sqrt{\frac{1}{2}} = 0.707$. This predicts a sol-to-gel transition will occur when a conversion of $A$ groups reaches 0.707. This is consistent with the appearance of a clear rubbery plateau for PPG0.75/0.8/0.85/1 (Fig. 4a) where the conversion of amine groups is above the critical value of 0.707 (0.75/0.8/0.85/1). PPG0.7/0.5/0.3/0 samples have the conversion of amine groups below the critical value, and these systems form longer but finite chains, consistent with the lack of rubbery plateau in the linear viscoelastic spectra. In this regime, the length of the formed chain will increase as more amine groups convert to crosslinks, and therefore the terminal flow regime will shift to a lower frequency when the crosslinker concentration increases. It is impressive that the power law $G^* \sim \omega^{1/3}$ for PPG0.7 extends for ~6 decades (Figs. 4a and 4b). This result is consistent with the viscoelasticity of randomly branched polyester samples near the sol-to-gel transition, where the complex shear modulus also showed a power law dependence with an exponent 0.33±0.05 [43].

For PPG0.75/0.8/0.85, one may find that the rubbery plateau ends with a drop in the real part of shear modulus $G'$ at lower frequencies (Fig. 4a), indicating rearrangement of imine bonds. It is fundamentally different from the networks with permanent crosslinks showing an infinite rubbery plateau [44]. Such behavior is observed for PPG1 with no free amine groups left. This result reveals there is no imine bond exchange in the experimental window for PPG1 even at the highest measured temperatures. It is consistent with the recent work [31], demonstrating that the often proposed imine metathesis mechanism [45] is unlikely without catalysts. We also note that spectra in PPG0.75/0.8/0.85 samples below the rubbery plateau do not obey $G'(\omega) \propto \omega^2$ and $G''(\omega) \propto \omega$, therefore these systems do not reach terminal flow in the measured frequency range.

Furthermore, an increase of the rubbery plateau modulus ($G_N$) with temperature was observed for each sample above gelation (Fig. 4c), which is consistent with the entropic elasticity of associative networks with fixed concentrations of cross-links [20]. Interestingly, the obtained $G_N$ values of PPG1 are slightly lower than the predictions of the phantom network model $G_N = \frac{f-2}{f} \times \frac{\rho RT}{M}$ (red line) [46,47], where $f$ is the cross-link functionality equal to 3 in our case, $\rho$ is density assumed to be 1.00 g/cm$^3$, and $M$ is the molecular weight of PPG-NH$_2$ chain equal to 4000 g/mol). The lower $G_N$ values are most likely caused by the formation of loops between PPG-NH$_2$ and BTA, instead of creating an ideal network. In addition, at a fixed temperature, $G_N$ decreases as crosslinker concentration decreases (Fig. 4d). Linear extrapolation of $G_N$ to the zero value (Fig. 4d) suggests the gelation point to be ~0.74, in good agreement with that predicted from the classical mean-field theory for permanent crosslinked networks.

We analyzed the temperature dependence of bond rearrangement time $\tau_r$ (the dashed line) for PPG0.85/0.8/0.75 using the traditional Arrhenius law (Eq. 3) (Fig. 3a). The determined apparent activation energies are 0.36, 0.46, and 0.44 eV (*i.e.*, 35, 44, and 42 kJ mol$^{-1}$) for PPG0.85/0.8/0.75, respectively, which are comparable to those previously reported in [31,48] for imine bonds. It is surprising that despite these relatively low activation energies, $\tau_r$ is extremely long (*i.e.*, 1-100s) even at very high temperatures (Fig. 3a). We ascribe this feature to the large pre-factor $A$: 10$^{-2.3}$, 10$^{-4.1}$, and 10$^{-4.5}$ s for PPG0.85/0.8/0.75, respectively. These values are ~8-10 orders of magnitude higher than the typical value of the molecular rearrangement pre-factor $\tau_0$ ~ 10$^{-14}$-10$^{-12}$ s [33,34].

Recently, this abnormally large pre-factor was ascribed to the often-overlooked chemical steric factor $p$ [32,49], which quantifies the probability that reactant molecules will collide with the correct orientation to form a chemical bond. Then, Eq. (3) was modified to account for the steric factor [32]:

$$\tau_r(T) = \frac{\tau_0}{p} \exp\left(\frac{E_a}{k_B T}\right) \quad (4)$$

It should be emphasized that Eq. (4) predicts the regime when segmental relaxation time is much shorter compared to the bond rearrangement time, and the chemical reaction/steric factor becomes the rate-limiting parameter. According to the activated complex theory [50], the steric factor $p$ can be described as:

$$p = \frac{\Omega^*}{\Omega} = \exp\left(\frac{S^* - S}{k_B}\right) = \exp\left(\frac{\Delta S}{k_B}\right) \quad (5)$$

where the quantities $\Omega^*$ and $\Omega$ describe the number of states accessible at activated complex and ground states, and $S^* = k_B \ln \Omega^*$ and $S = k_B \ln \Omega$ are the entropies in the activated complex and ground states, respectively. This equation demonstrates clearly that the chemical steric factor originates from the activation entropy for these reactions. Substituting Eq. 5 to Eq. 4 and considering $\tau_0 = \frac{h}{k_B T}$ (here $h$ is Planck's constant), $\tau_r(T)$ can be rewritten as follows:

$$\tau_r(T) = \frac{h}{k_B T} \exp\left(-\frac{\Delta S}{k_B}\right) \exp\left(\frac{E_a}{k_B T}\right) \quad (6)$$

When comparing Eq. 6 with Eq. 3, the pre-factor $A$ in Eq. 3 is described as follows at temperatures from 300 K to 400 K:

$$A = \frac{h}{k_B T} \exp\left(-\frac{\Delta S}{k_B}\right) \approx 10^{-13} \exp\left(-\frac{\Delta S}{k_B}\right) \quad (7)$$

According to Eq. 7, the extremely large pre-factor for the bond rearrangement process in studied vitrimers is due to the large negative activation entropy for the bond exchange. We ascribe this negative entropy term to very precise atomic positions and bond angles arrangements required for successful dynamic covalent bond switching. In simple terms, it requires a unique transition state relative to the many states accessible for a free $NH_2$ group in its ground state, leading to large negative $\Delta S$ (Eq. 5). By fitting $\tau_r(T)$ to Eq. 6, we obtained the activation entropy $\Delta S$ to be $-2.2\times10^{-3}$, $-1.8\times10^{-3}$, and $-1.8\times10^{-3}$ eV K$^{-1}$ (*i.e.*, -210, -176, and -169 J mol$^{-1}$ K$^{-1}$) for PPG0.85/0.8/0.75, respectively. Furthermore, based on the reported pre-factor $A$ in literature for other vitrimers, we estimated the activation entropies varying from $-9.3\times10^{-4}$ to $-2.2\times10^{-3}$ eV K$^{-1}$ (*i.e.*, -90 to -210 J mol$^{-1}$ K$^{-1}$) for imine bond based vitrimers and from $-1.2\times10^{-3}$ to $-2.3\times10^{-3}$ eV K$^{-1}$ (*i.e.*, -120 to -220 J mol$^{-1}$ K$^{-1}$) for boronic/boric ester bond based vitrimers (Fig. 3b). Remarkably, these values of the activation entropies are much higher than those reported for hydrogen bonded urea-terminated telechelic PDMS systems with $\Delta S$ varying from $-6.2\times10^{-5}$ to $-3.2\times10^{-4}$ eV K$^{-1}$ (*i.e.*, -6 to -31 J mol$^{-1}$ K$^{-1}$) (Fig. 3b), even though their bond dissociation energy 0.31 eV (~30 kJ mol$^{-1}$) is comparable to that of imine and boronic/boric ester bond exchange [51]. This clearly demonstrates the general and critical role of activation entropy in slowing down the associative bond exchange process in vitrimers. Apparently, perfect atomic alignment required for the dynamic covalent bond switching slows down the bond rearrangement significantly even if the enthalpic energy barrier is relatively low. This large negative entropic term should be explicitly considered in any modelling of vitrimers. Furthermore, using Eq. 5, we can estimate the ratio of the number of states in activated complex to that in the grounded state, i.e., the chemical steric factor $p$. The estimated values of $\frac{\Omega^*}{\Omega}$ are ~$10^{-11}$ -$10^{-5}$ and ~$10^{-12}$ -$10^{-7}$ for imine-bonded and boric/boronic ester vitrimers, which are a million to trillion times lower than those in H-bonded associative polymers where $\frac{\Omega^*}{\Omega}$ ~0.01-1. This clearly demonstrates the exceptionally limited numbers of accessible activated state in vitrimers required for the bond exchange process, resulting in a significantly prolonged bond rearrangement time. A detailed calculation of the associative bond switching through density functional theory might be crucial in providing further microscopic perspective of this entropic term, specific for each type of dynamic covalent bond chemistry.

In summary, we studied the relaxation dynamics and viscoelastic behavior of a series of model imine bond-based vitrimers. Our study focused on the effects of crosslink density and temperature on the segmental relaxation, chain dynamics, and terminal flow/bond rearrangement processes. We found that while crosslink density had little impact on the temperature-dependent segmental and chain dynamics, it significantly influenced the linear viscoelasticity of studied samples. Specifically, increasing the crosslink density led to a clear sol-gel transition at the gelation point, which agrees well with predictions from classical sol-gel theory. Below gelation, the viscoelastic behavior followed Rouse dynamics. Near the gel point, we observed a broad intermediate power-law regime in the complex shear modulus covering 6 orders in frequency between the Rouse and terminal flow regimes. Above the gelation point, a rubbery plateau appears which is terminated at lower frequencies by the covalent bond rearrangement process. This process is not observed in the system with no free $NH_2$ groups, confirming that metathesis is not a viable mechanism for imine bonds rearrangements. Surprisingly, analysis of the temperature dependence of the bond rearrangement times revealed a large negative activation entropy associated with this process, suggesting that entropic barriers play a major role in slowing associative bond exchange. We suggest that the very specific atomic and bond angles arrangement required for the associative bond exchange is the major reason for the surprisingly large negative activation entropy of this process. This explains the frequently obtained long bond rearrangement times despite the relatively low enthalpic barrier for bond exchange in vitrimers at temperatures far above $T_g$.

**Acknowledgements**

We thank Jeff Foster and David Simmons for many helpful discussions. This work is partially supported by the NSF Polymer program (DMR- 2515834).

# SUPPLEMENTARY MATERIALS

Unexpectedly large entropic barrier controls bond rearrangements in vitrimers

*Shinian Cheng[1*], Lilliana Rey[1], Beibei Yao[2], Ivan Popov[3], Alexei P. Sokolov[1,2*]*

[1]Department of Chemistry, University of Tennessee, Knoxville, Tennessee 37996, USA

[2]Chemical Sciences Division, Oak Ridge National Laboratory, Oak Ridge, Tennessee 37831, USA

[3]University of Tennessee-Oak Ridge Innovation Institute, University of Tennessee, Knoxville, Tennessee 37996, USA

*Author Correspondence should be addressed to scheng17@utk.edu and sokolov@utk.edu

## 1. Methods

**Materials and synthesis of model systems** Poly(propylene glycol) bis(2-aminopropyl ether) (PPG-NH$_2$, $M_n$=4000 g/mol) and Tetrahydrofuran (THF, ≥ 99.9%, inhibitor-free) were purchased from Sigma-Aldrich. Benzene-1,3,5-tricarbaldehyde (BTA, >98.0%) was purchased from TCI America. All samples were used as received. For the synthesis of PPG-based vitrimers with imine bonds, 3 g of PPG-NH$_2$ were dissolved in 15mL of THF. This solution under stirring was added dropwise to 10 mL of BTA solutions in THF whose concentrations were adjusted depending on the desired degree of functionalization. This mixture was reacted at 40 °C under nitrogen protection for 18 h. After having most of the solvent evaporated using rotary evaporation, the solution was poured into a Teflon disk and moved to a vacuum oven for further evaporation at 40 °C for 24 h, and then to dry at 85 °C for 48 h.

**Broadband Dielectric Spectroscopy (BDS)** Dielectric spectra in the frequency range between $10^{-2}$ and $10^6$ Hz were measured using a Novocontrol system. This apparatus included an Alpha-A impedance analyzer and a Quatro Cryosystem temperature control unit. The samples were placed into a parallel plate dielectric cell made of sapphire and invar steel with an electrode diameter of 10 mm and a capacitance of ~3.45 pF with an electrode separation of ~0.25 mm. All the spectra were measured on both cooling and heating in a temperature range from 203 K to 333K. The samples were equilibrated for 10 min at each temperature to reach thermal stabilization within 0.2 K. An electric voltage of 1V was applied for all measurements. Before BDS measurements, the sample in the dielectric cell was dried in a vacuum oven at 90 °C for 24h.

**Small amplitude oscillatory shear rheology (SAOS)** Linear viscoelastic properties of all samples were explored through small amplitude oscillatory shear rheology measurements using a strain-controlled rheometer (AR2000ex, TA Instruments) in the angular frequency range of $10^{-1}$-$10^2$ rad/s. A parallel plate geometry with diameters of 4 and 8 mm was employed, depending on the magnitude of the shear modulus. The sample thickness at room temperature was around 0.9 mm. Before rheology measurements, the sample was dried in the rheometer at 423K for 2 h under $N_2$ flow.

**Raman spectroscopy, small-angle X-ray scattering (SAXS), and differential scanning calorimetry (DSC)** The chemical structures of the synthesized samples were verified using a T64000 Raman spectrometer from Horiba Jobin Yvon in the single pass mode. The experiments were performed using a backscattering geometry and laser with the wavelength 633 nm at room temperature. Prior to the measurements the spectrometer was calibrated using a silicon wafer. The absence of nanoscale phase separation was confirmed by small angle X-ray scattering measurements using a Xenocs Xeuss 3.0 instrument, and the glass transition temperature of the samples were evaluated with DSC2500 (TA instruments).

2. **Normalized Raman spectra**

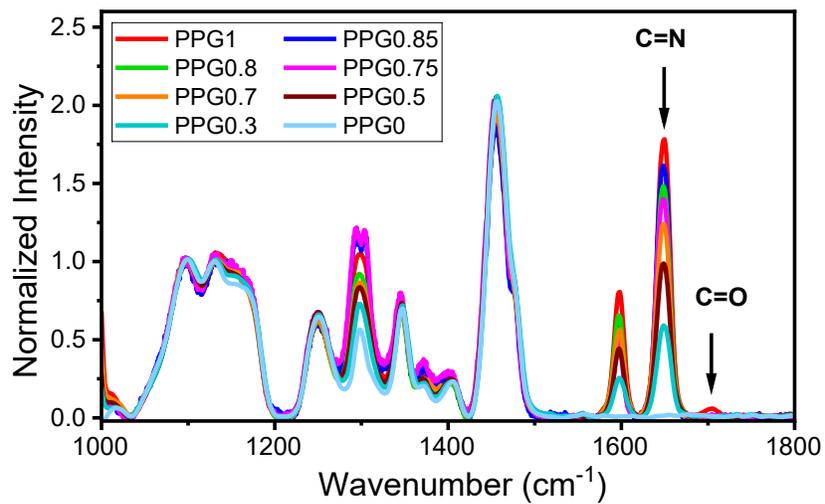

Figure S1: Raman spectra of samples studied normalized to the peak of C-O bond at 1099 cm$^{-1}$.

## 3. DSC results

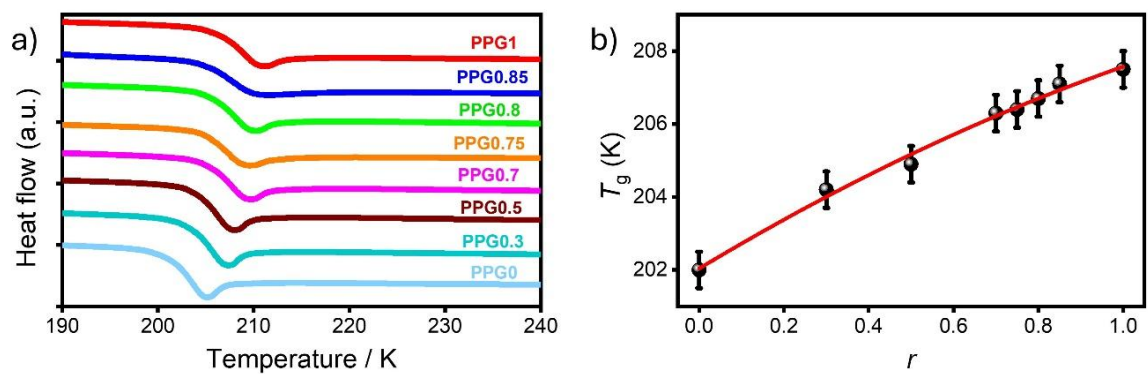

Figure S2: a) DSC heating results of samples studied at a 10K/min rate; 6) The glass transition temperature $T_g$ as a function of crosslink density $r$.

### 4. BDS spectra analysis

The dielectric spectra of samples were analyzed using two Havriliak-Negami (HN) functions corresponding to a normal mode and a segmental relaxation mode:

$$\varepsilon^*(\omega) = \varepsilon_\infty + \frac{\Delta\varepsilon_n}{\left(1+(i\omega\tau_{HN,n})^{\beta_n}\right)^{\gamma_n}} + \frac{\Delta\varepsilon_s}{\left(1+(i\omega\tau_{HN,s})^{\beta_s}\right)^{\gamma_s}} + \frac{\sigma_{DC}}{i\omega\varepsilon_0} + A\omega^{-b} \quad (S1)$$

where $i$ is the imaginary unit, $\varepsilon^*$ is the complex permittivity, $\varepsilon_\infty$ and $\varepsilon_0$ are the dielectric constants at the infinite high frequency and the vacuum permittivity, $\omega$ is the angular frequency, $\tau_{HN,n}$ and $\tau_{HN,s}$ are the characteristic HN time of the normal mode and segmental relaxation, $\Delta\varepsilon_n$ and $\Delta\varepsilon_s$ are the dielectric amplitudes of the normal mode and the segmental relaxation, $\sigma_{DC}$ is the dc-conductivity, $\beta_s$ and $\gamma_s$ are the shape parameters of the segmental relaxation, $\beta_n$ and $\gamma_n$ are the shape parameters of the normal mode, and $A$ and $b$ are constants. The characteristic relaxation time of the normal mode ($\tau_n$) and the segmental relaxation mode ($\tau_s$) can be obtained from the characteristic HN time:

$$\tau_j = \tau_{HN,j}\left[\sin\frac{\beta_j\pi}{2+2\gamma_j}\right]^{-1/\beta_j}\left[\sin\frac{\beta_j\gamma_j\pi}{2+2\gamma_j}\right]^{1/\beta_j}, \text{ with } j=n \text{ or } s$$

The representative spectra of $\varepsilon''(\omega)$ of samples studied are shown in Figure S3 below.

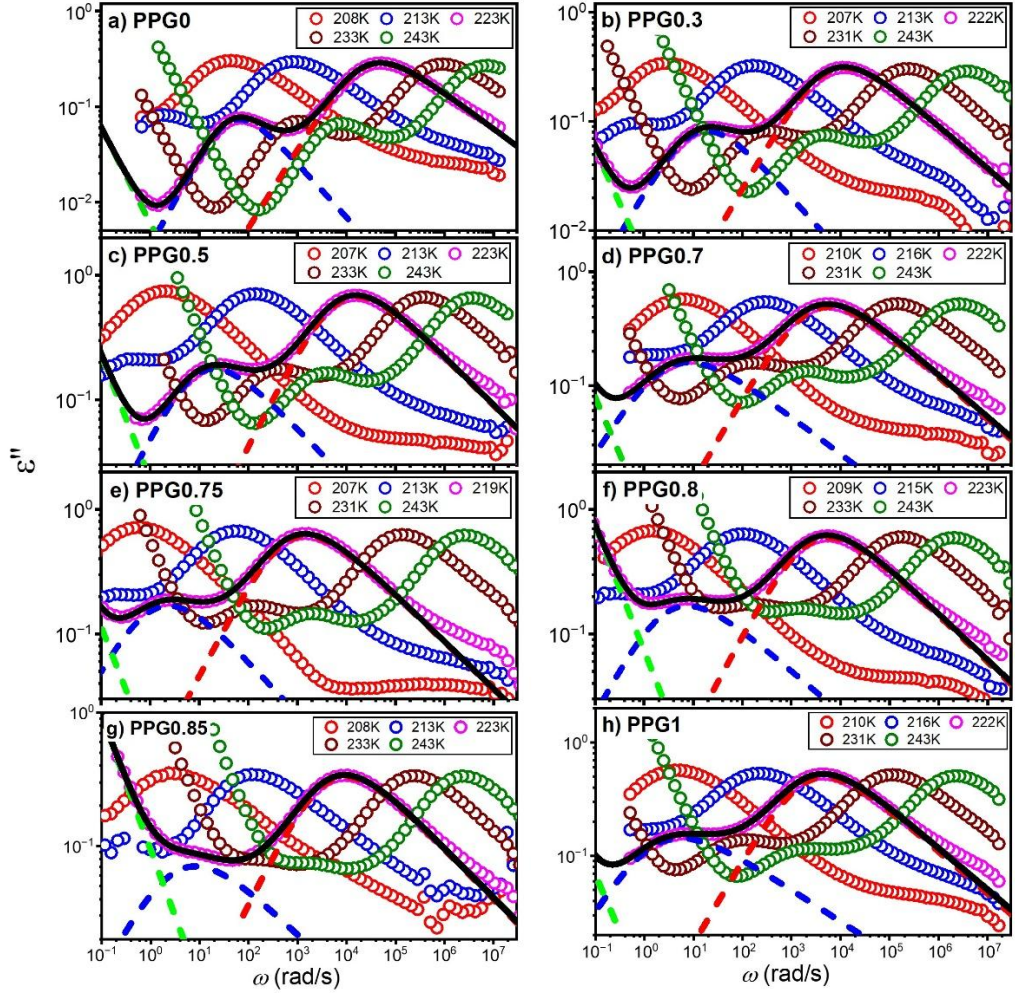

Figure S3: Representative spectra of ε″ (ω) obtained from BDS for studied samples. The lines in each figure show the representative fit of the spectra to Eq. S1. The green, blue, and red dashed lines present the dc-conductivity contribution, the normal mode, and segmental relaxation mode, respectively. The black solid lines present the sum of these three modes, showing good agreement with the experimental data.

## 5. Time-temperature superposition of rheology spectra at $T_{\text{ref}}$=230K

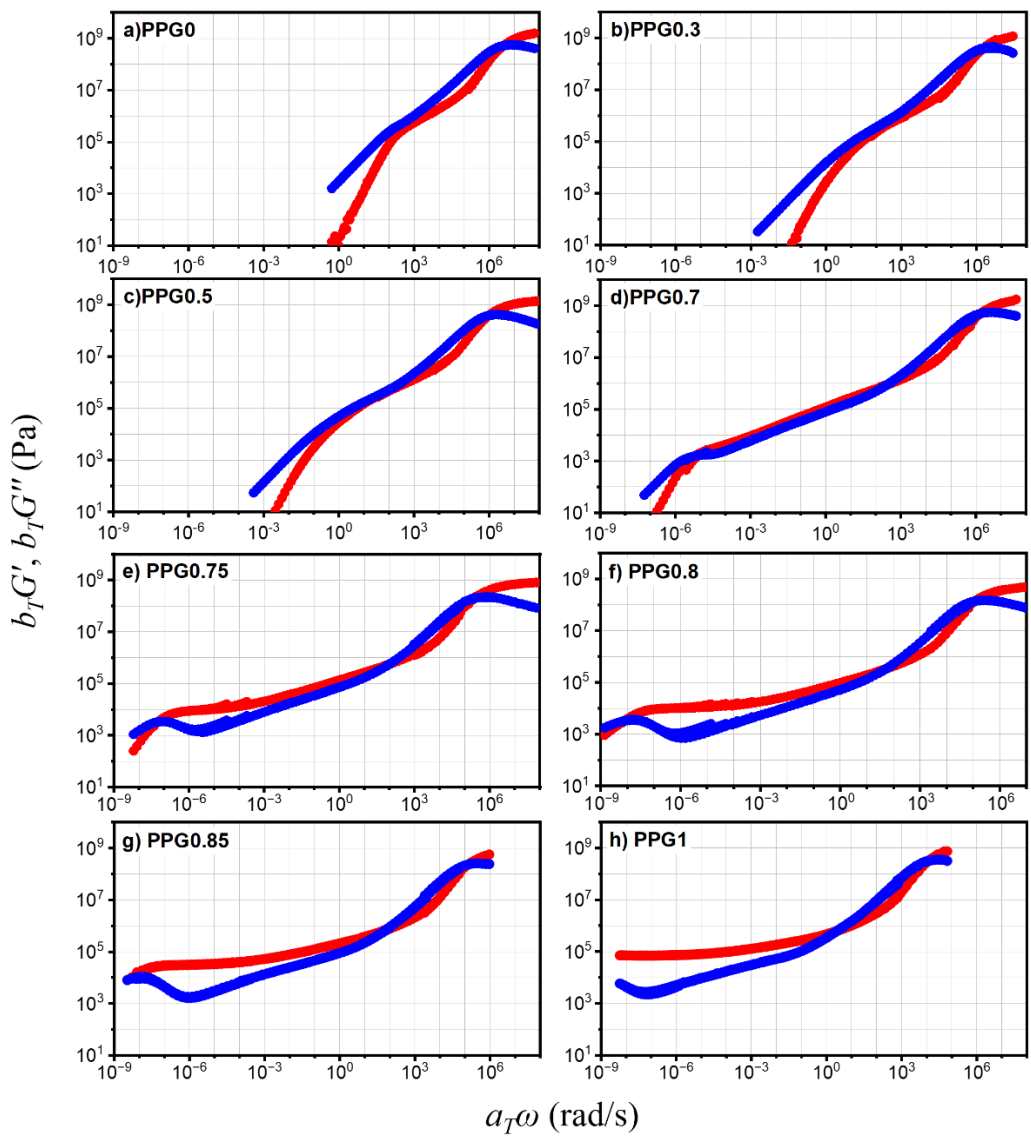

Figure S4: Time-temperature superposition master curves of $G'$ (red) and $G''$ (blue).